\documentclass[preprint2]{aastex}

\usepackage{savesym}
\usepackage{amsmath}
\savesymbol{iint}
\usepackage{txfonts}
\restoresymbol{TXF}{iint}
\usepackage{longtable}
\usepackage{rotating}
\usepackage{natbib}
\usepackage{graphicx}
\usepackage{graphics}
\usepackage{psfrag}
\usepackage{bm}
\usepackage[
pdfauthor={derajan},
pdftitle={How to do this},
pdfstartview=XYZ,
bookmarks=true,
colorlinks=true,
linkcolor=blue,
urlcolor=blue,
citecolor=blue,
pdftex,
bookmarks=true,
linktocpage=true, 
hyperindex=true
]{hyperref}
\usepackage{amssymb}
\bibpunct{(}{)}{;}{a}{}{,}

\def\kms{$\mathrm{km\,s}^{-1}$}

\def\gs{$\mathrm{g\,s}^{-1}$}
\def\Rj{\ensuremath{R_{\rm Jup}}}
\def\Mj{\ensuremath{M_{\rm Jup}}}

\def\Rpl{\ensuremath{R_{\rm pl}}}
\def\Mpl{\ensuremath{M_{\rm pl}}}

\def\apjl{ApJL}
\def\arep{ARep}
\def\solphys{SoPh}
\def\nat{Natur}
\def\icarus{Icar}

\newcommand{\Lp}[1]{${\mathrm{L_{#1}}}$}

\shorttitle{The Influence of CMEs on the Mass-loss Rates of Hot-Jupiters}
\shortauthors{Cherenkov et al.}

\begin{document}


\title{The Influence of Coronal Mass Ejections on the Mass-loss Rates of Hot-Jupiters}

\author{A. Cherenkov and D. Bisikalo\altaffilmark{1}}
\affil{Institute of Astronomy of the Russian Academy of Sciences, 48 Pyatnitskaya St. 119017, Moscow, Russian Federation}

\author{L. Fossati and C. M\"ostl}
\affil{Space Research Institute, Austrian Academy of Sciences, Schmiedlstrasse 6, A-8042 Graz, Austria}

\altaffiltext{1}{bisikalo@inasan.ru}

\begin{abstract}
Hot-Jupiters are subject to extreme radiation and plasma flows coming from their host stars. Past ultraviolet \textit{Hubble Space Telescope} observations, supported by hydrodynamic models, confirmed that these factors lead to the formation of an extended envelope, part of which lies beyond the Roche lobe. We use gas-dynamic simulations to study the impact of time variations in the parameters of the stellar wind, namely that of coronal mass ejections (CMEs), on the envelope of the typical hot-Jupiter HD\,209458b. We consider three CMEs characterized by different velocities and densities, taking their parameters from typical CMEs observed for the Sun. The perturbations in the ram-pressure of the stellar wind during the passage of each CME tear off most of the envelope that is located beyond the Roche lobe. This leads to a substantial increase of the mass-loss rates during the interaction with the CME. We find that the mass lost by the planet during the whole crossing of a CME is of ${\approx}10^{15}$\,g, regardless of the CME taken into consideration. We also find that over the course of 1\,Gyr, the mass lost by the planet because of CME impacts is comparable to that lost because of high-energy stellar irradiation.
\end{abstract}
%



\section{Introduction}\label{sec:intro}
The discovery, to date, of over 200 close-in giant planets orbiting their host stars at distances of less than 0.1\,au (hot-Jupiters) is one of the biggest surprises in the last few decades of planetary astronomy. \textit{Hubble Space Telescope} (\textit{HST}) ultraviolet (UV) observations of the hot-Jupiter HD\,209458b, conducted during primary transit with the STIS spectrograph, revealed transit depths at Ly$\alpha$ wavelengths of 9\%--15\%, compared to $\approx$2\% at optical bands \citep{Vidal-Madjar-2003,Ben-Jaffel-2007}. Rather deep transits have also been detected for far-UV lines of C, O, and Si \citep{Vidal-Madjar-2004,Ben-Jaffel-2009,Linsky-2010}. Large transit depths at the wavelengths of resonance lines of abundant elements have also been observed for the hot-Jupiters HD\,189733b \citep{lecavelier2012} and WASP-12b \citep{Fossati-et-al:2010a,haswell2012}, and for the warm Neptune-mass planet GJ\,436\,b \citep{ehrenreich2015}.

These observations indicate the presence of atmospheric material beyond the Roche lobe of these planets. The energy deposited by the stellar high-energy radiation onto the upper atmosphere of a planet increases the temperature of the thermosphere that then expands. As a consequence, the upper part of the atmosphere (exosphere) may move beyond the planet's Roche lobe, hence escaping to space \citep[e.g.,][]{lammer2003,yelle2004,koskinen2007,koskinen2013a,koskinen2013b,bourrier2013,Kislyakova-14}.

Because of their proximity to the host star, hot-Jupiters are subject to extreme irradiation and the gravitational influence of the host star. This leads, for example, to significant changes in the upper atmospheres of planets (due to Roche lobe overflow) and to the presence of a large dynamic pressure of the stellar wind, which in turn favors the formation of strong bow shocks, provided that the orbital velocity exceeds the local sound speed of the stellar wind \citep{vidotto2010,Bisikalo2013a}. These effects have a substantial influence on the interaction between the expanding planetary atmosphere and the stellar wind.

\citet{Bisikalo2013a, Bisikalo-2013b-Rus} presented the results of 3D hydrodynamic simulations aiming to analyze the interaction between the expanding atmosphere of hot-Jupiters and the stellar wind. They showed that, for typical hot-Jupiters orbiting their host stars with supersonic velocities, the interaction of the planetary atmosphere with the stellar wind has a significant influence on the geometry and physical parameters of the atmosphere. In particular, the ``type'' of planetary atmosphere (i.e., closed, quasi-closed, or open; see below) depends on the position of the head-on collision point (where the dynamic pressure of the stellar wind equals the pressure of the expanding planetary atmosphere). The envelope of planets for which the head-on collision point is located inside the Roche lobe is almost spherical (closed) and only slightly distorted by the gravitational influence of the star and stellar wind. If the head-on collision point is instead located beyond the Roche lobe, the atmosphere starts to outflow through the \Lp1 and \Lp2 Lagrangian points. This leads to the formation of an extended asymmetric envelope. If the dynamic pressure of the stellar wind is strong enough to stop the outflow through the inner Lagrangian point \Lp1, a quasi-closed stationary envelope forms. If instead the stellar wind is not capable of stopping the outflow from \Lp1, the planet will develop an open envelope. Detailed simulations of these three cases have shown that the mass-loss rates of a quasi-closed envelope ($\dot{M} = 3 \times 10^{9}$\,\gs) are similar to those of a closed envelope ($\dot{M} = 2 \times 10^{9}$\,\gs), while for an open envelope the mass-loss rates are about a factor of 10 higher \citep[$\dot{M} = 3 \times 10^{10}$\,\gs;][]{Cherenkov-14}. This implies that quasi-closed envelopes may be stationary.

Since hot-Jupiters have extended hydrogen envelopes, the matter outside the Roche lobe is weakly gravitationally bound to the planet, so perturbations in the stellar wind can lead to changes in the planetary atmospheric structure and mass-loss rate. Even inactive late-type stars, such as the Sun, possess a wind that may undergo drastic temporal variations. The major perturbations to the solar wind are due to large ejections of matter from the solar corona, so-called coronal mass ejections (CMEs). In the case of the Sun, CMEs are characterized by a mass of plasma ejected into the interplanetary medium of approximately $10^{15}$\,g, an average total energy of about $10^{31}$\,erg, and ejection velocities that vary from about 20--3000\,km\,s$^{-1}$, with averages of the order of 500\,km\,s$^{-1}$ \citep{vourlidas2010,webb2012}. CMEs expand with supersonic velocity, so the propagation of the ejection is accompanied by the formation of a shock. Both the average speed and the frequency of solar CMEs vary with the solar activity cycle. Note that even for the Sun, the frequency of CMEs is rather high, going from about 0.5 per day during solar minimum to up to about 4 per day during solar maximum \citep{webb2012}. Averaging over a full solar cycle, CMEs impact Earth about twice per month \citep{richardson2010}. It is therefore important to evaluate the impact of CMEs on the atmosphere of hot-Jupiters, particularly on the mass-loss rates, to estimate their influence on planetary evolution.

\citet{khodachenko2007} and \citet{lammer2007} studied the effects of CMEs erupting from low-mass stars on the secondary atmospheres of Earth-like planets in the habitable zone. They concluded that weakly magnetized planets may lose hundreds of bars of atmospheric pressure under the action of CMEs. \citet{kay2016} studied the deflection of CMEs from planetary magnetic fields, concluding that planets orbiting M-dwarfs require magnetic fields between tens and hundreds of Gauss to be able to avoid the eroding effect of CMEs on the atmosphere. For hot-Jupiters orbiting Sun-like stars, the required magnetic fields are of the order of a few tens of Gauss \citep{kay2016}, though this is still much larger than what was estimated for HD\,209458b \citep{Kislyakova-14}. This shows that CMEs may play an important role in the atmospheric evolution of planets, particularly if orbiting close to the host star, which is the case for hot-Jupiters or planets in the habitable zone of low-mass stars.

We present results obtained from investigating the effects of CMEs on the flow structure of the envelope of a typical hot-Jupiter. \citet{Cherenkov-et-al-15} presented preliminary results obtained from modeling the interaction between CMEs and the envelope of a hot-Jupiter similar to HD\,209458b. That work was carried out using the same modeling tools adopted here, but considering a single CME with a simplified structure and for which the parameters were chosen such that the relative changes in density and temperature during the CME at the distance of 0.05\,au were the same as for the CME that hit Earth on 1998 April 12 \citep{Farrell_2012}. In this work, we build upon the results of \citet{Cherenkov-et-al-15} using proper CME parameters for the typical orbital separation of hot-Jupiters and by considering that the CME parameters (velocity, duration, and density) can be very different and variable in time. In this work, we consider a HD\,209458b-like hot-Jupiter being hit by three CMEs with different propagation velocities and density variations.

This paper is organized as follows. In Section~\ref{sec:model} we present the physical and computational model. In Section~\ref{sec:cme-params} we discuss the adopted CME parameters. In Section~\ref{sec:cmeresult} we present the results, which are then discussed in Section~\ref{sec:discussion}. In Section~\ref{sec:conclusion} we draw our conclusions.
\section{The Hydrodynamic Model}\label{sec:model}
As a typical hot-Jupiter, we consider here the transiting exoplanet HD\,209458b, which has a radius \Rpl\ of 1.38\,\Rj\ and a mass \Mpl\ of 0.69\,\Mj. The planet orbits a main-sequence G0 star at a distance of 0.04747\,au, corresponding to an orbital period of 3.52472\,days\footnote{http://exoplanet.eu/ -- \citet{schneider2011}}. We set the temperature and density of the atmosphere at \Rpl\ equal to the values given in \citet{Bisikalo-2013b-Rus}, where those parameters were taken from \citet{koskinen2013a}. Those values lead to a quasi-closed solution, with about 1.11$\times$10$^{15}$\,g of gas lying between the Roche lobe and the contact discontinuity (note that the mass of gas lying between \Rpl\ and the Roche lobe is about 1.7$\times$10$^{17}$\,g). In particular, we adopted at \Rpl\ a temperature $T$ of 7.5$\times$10$^3$\,K and a gas density $n$ of 10$^{11}$\,cm$^{-3}$. Following \citet{Bisikalo-2013b-Rus}, we adopted the parameters of the stationary (i.e., without CME) stellar wind equal to those of the solar wind at the planet's orbital separation, corresponding to a temperature $T_w$ of 7.3$\times$10$^5$\,K, density $n_w$ of 10$^4$\,cm$^{-3}$, and radial velocity $v_w$ of 100\,\kms\ \citep{Withbroe-1988}. With these parameters, the stationary stellar wind is slightly subsonic; however, given the supersonic orbital motion of the planet (Mach number $M$ equal to 1.4), the total velocity of the planet relative to the stationary stellar wind is supersonic ($M$\,=\,1.75).

To model the system, we use the Roche approach, within which the star and planet are represented by point masses, moving in circular orbits around the system's center of mass. Since the mass of the gaseous envelope is much smaller than the planetary mass, we ignore the self-gravity of the gas. With this set-up, the Roche potential has five libration points, the Lagrangian points, where the gradient is equal to zero. The equipotential, going through the inner Lagrangian point \Lp1, encloses two contiguous volumes, known as critical surfaces or Roche lobes.

The numerical model adopted to study the flow structure in the planetary envelope is presented in \citet{Bisikalo-2013b-Rus}. We used a three-dimensional system of equations to describe the gravitational gas dynamics, closed by the equation of state of an ideal, monoatomic gas

\noindent
\begin{eqnarray}
   &\dfrac{\partial \rho}{\partial t} +
   \nabla \cdot (\rho \bm{u}) = 0, \\
  &\dfrac{\partial }{\partial t} (\rho \bm{u})+
 \nabla \cdot  \big[ \rho \bm{u}  \otimes \bm{u} + p \, I\big] =  - \rho \cdot \mathrm{grad} \, \Phi + \rho \cdot  \bm{a_C}, \\
 &\dfrac{\partial \varepsilon}{\partial t} +
 \nabla \cdot \big[ \bm{u} \, (\varepsilon + p) \big] =  - \rho \bm{u} \cdot \mathrm{grad} \, \Phi,
\end{eqnarray}

\noindent
where $\rho$ is the gas density, $\bm{u}$ is the gas velocity, $p$ is the pressure, $\varepsilon = \rho \epsilon + \rho \bm{u}^2/2$ is the total energy density, $\bm{a_C} = - 2 \, [ \bm{\Omega} \times \bm{u}]$ is the Coriolis acceleration, $\Omega$ is the angular velocity of the star--planet reference system rotation, and $\Phi$ is the gravitational Roche potential.


In our computations, we neglected the stellar radiation pressure and the magnetic field of the planet. The first assumption is valid since at the adopted temperature, the atmospheric hydrogen is mostly ionized and hence the influence of the stellar radiation is small. The second assumption is supported by the estimate that the magnetic moment of HD\,209458b should be less than 0.1 times that of Jupiter \citep{Kislyakova-14}. By assuming such a weak magnetic field, the gas-dynamic escape plays a dominant role for HD\,209458b, because the lower atmosphere is dominated by neutral gas, while at higher altitudes the gas-dynamic is driven by the thermal pressure \citep{Arakcheev-et-al-17, Bisikalo-et-al-17}. We do not consider the magnetic field of the stellar wind, as its magnetic pressure is comparable to that of the dynamical pressure (in the stationary case and during CME passage), and it would not change the main results of this work. We also neglect the variations of the stellar wind chemical composition during the CME crossing.

The code solves the system of gas-dynamic equations in the Cartesian coordinate frame $(X,Y,Z)$ with the origin in the star's center. The $X$ axis is directed from the star to the planet, the $Z$ axis is aligned with the vector of rotation of the system $\bm{\Omega}$, and the $Y$ axis complements the right-handed system. The coordinate frame rotates along with the star--planet system. The non-uniform Cartesian grid has the size of 468$\times$468$\times$178 elements in the $X$, $Y$, and $Z$ dimensions, respectively. The physical size of the computational domain is 40$\times$40$\times$10\,\Rpl. The grid was locally refined close to the planet such that the size of each element at \Rpl\  ($\Delta x$ and $\Delta y \simeq 0.04$\,\Rpl) is smaller than one pressure scale height. The simulation grid has been set such that the scales of the major flow elements, located near the boundaries of the computational domain, are significantly larger than the size of the cells in this region ($\Delta x$ and $\Delta y \simeq 0.25$\,\Rpl). The mass-loss rate of the planetary envelope was computed as the difference between the fluxes of matter leaving and entering the system in a parallelepiped specified along the $X$, $Y$, and $Z$ axes, as large as 32$\times$32$\times$10\,\Rpl, containing the gaseous envelope. The choice of such a large volume is driven by the fact that we need to consider the entirety of atmospheric gas surrounding the planet. The computations were stopped after 1.7, 3.6, and 7.8 hr for the fast, medium, and slow CMEs (see Section~\ref{sec:cme-params}), respectively. After these times, the mass-loss rate remains constant at the value of the stationary wind.

In our simulations we use a TVD Roe--Osher numerical scheme. This explicit scheme of higher-order approximation possesses low numerical viscosity in the regions of smooth solution and does not smear out shock waves. This scheme, described in detail in \citet{BinaryStars:2002}, is based on the Roe scheme designed to solve the equations of gas dynamics. To avoid unphysical discontinuities, occurring in the unmodified Roe scheme, we apply the Einfeldt entropy fix \citep{Einfeldt-99}. The Osher TVD correction is then applied to increase the order of spatial approximation \citep{Chakravarthy-Osher}. The code is parallelized using the MPI library and the calculations were carried out with 324 processors of the Kurchatov Institute Supercomputer.
\section{The Adopted CME Parameters}\label{sec:cme-params}
%
\begin{figure}[h!]
\begin{center}
\includegraphics[width=\hsize,clip]{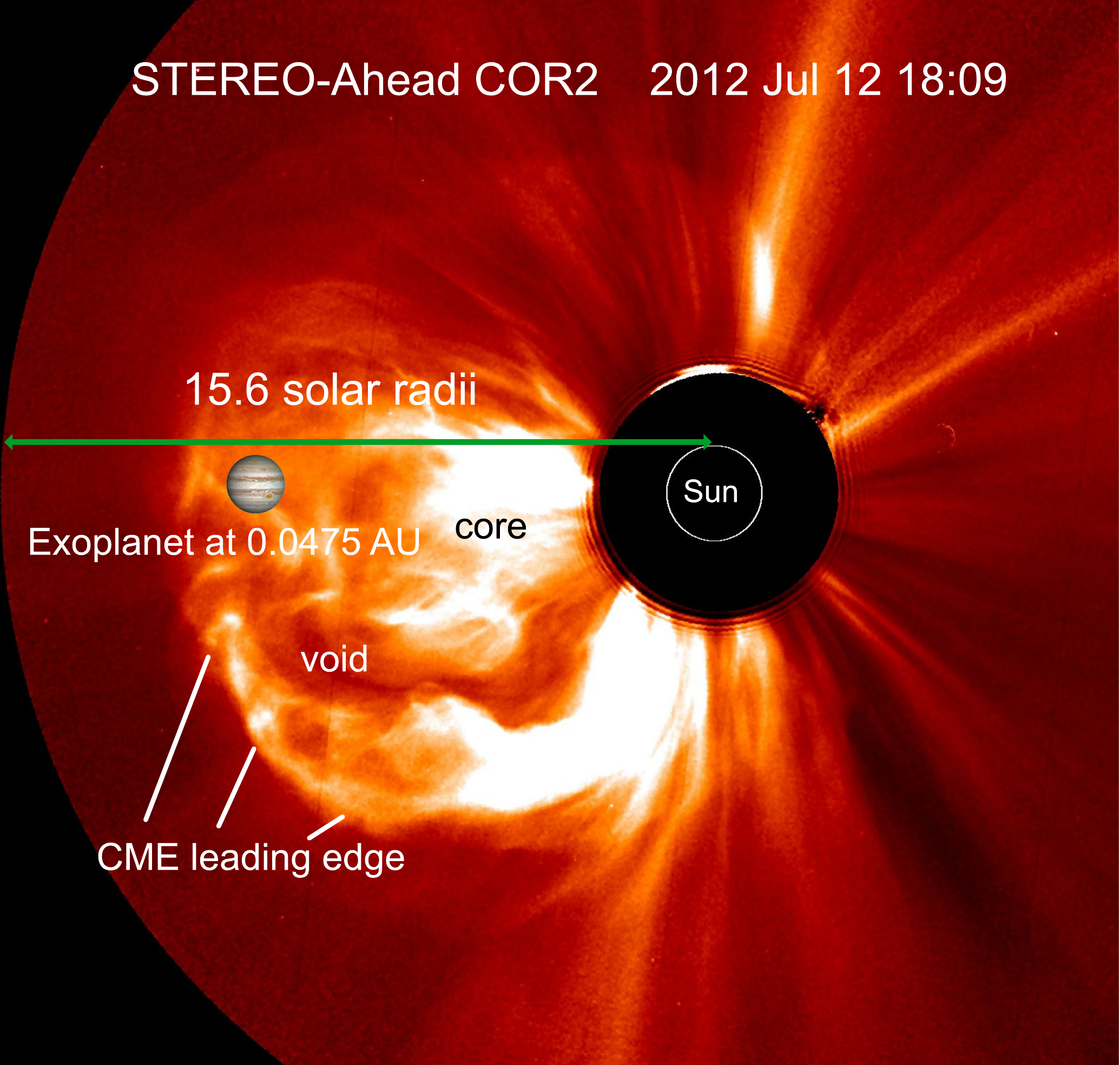}
\caption{Illustration of the star--planet--CME configuration under study. The figure indicates the position of the three parts of which a CME consists of, which are the leading edge, void, and core. The different intensities of these regions are related to the integrated density along the line of sight. Within the scale of this figure, the hot-Jupiter is located at the correct distance from the star, while in the scale of the original image of the solar CME the planet would be located well inside the coronagraph field of view of the STEREO/COR2 instrument, extending up to 0.073\,au. The size of the hot-Jupiter has been enhanced for better visibility. The star HD\,209458 has a radius of 1.148\,R$_\odot$.}
\label{picture:cor_cme} 
\end{center}
\end{figure}
\begin{figure}[h!]
\begin{center}
\includegraphics[width=\hsize,clip]{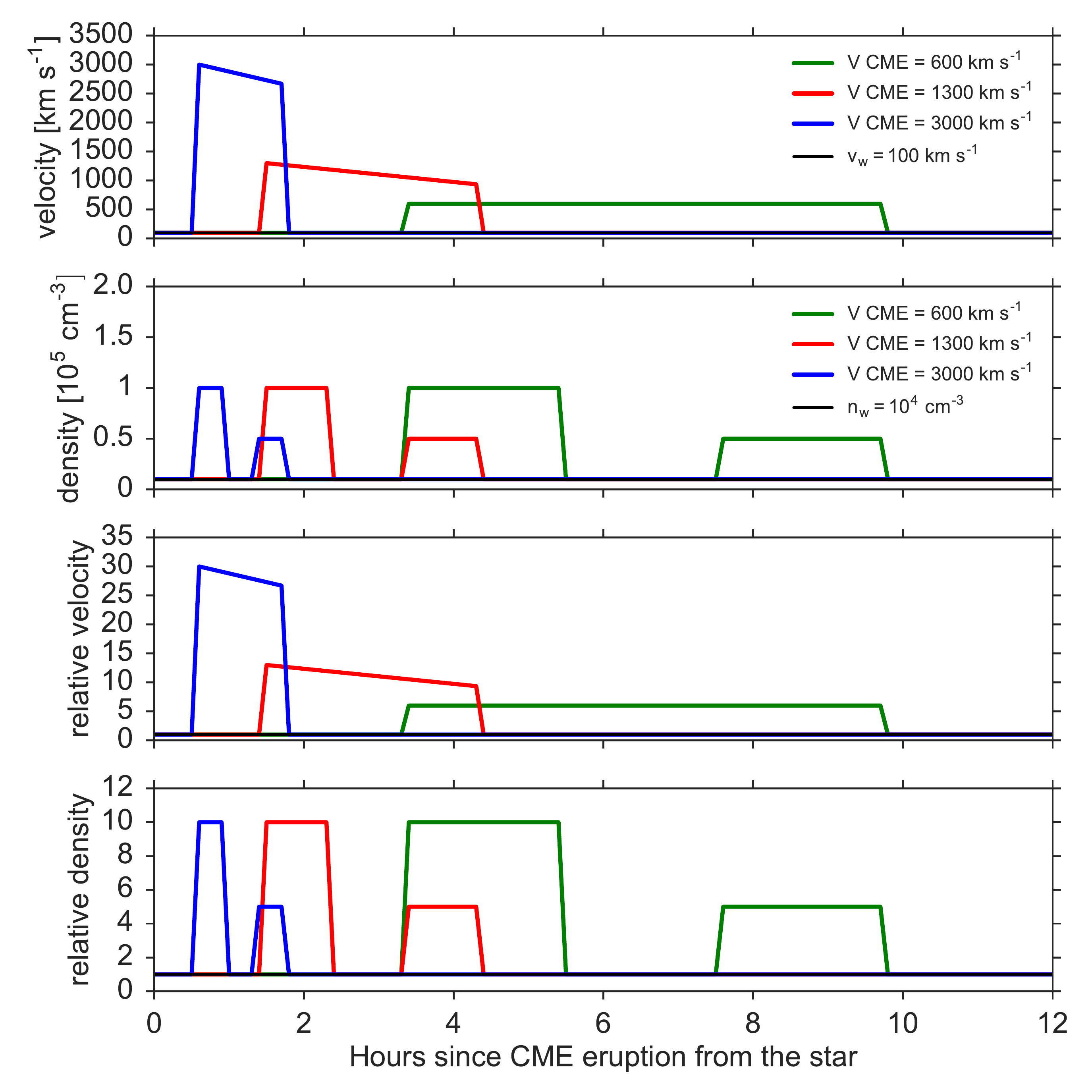}
\caption{Velocity and density profiles of the stellar wind during the slow (green), medium (red), and fast (blue) CMEs considered here as a function of time (in hours). The black horizontal line shows the level of the stationary stellar wind. The two bottom panels show the wind velocity and density relative to those of the stationary wind.}
\label{picture:wind} 
\end{center}
\end{figure}

In this work we consider CMEs with three different sets of parameters separated with respect to their speed of propagation: fast, medium, and slow.  The parameters are defined for a hot-Jupiter orbiting a Sun-like star at 0.05\,au, using empirical considerations derived from solar observations.

Figure~\ref{picture:cor_cme} shows an observation of a solar CME by the \emph{STEREO-Ahead} COR2 coronagraph instrument \citep{howard2008} on 2012 July 12. We have placed an image of Jupiter at the correct distance for HD\,209458b, which shows that coronagraphs can be used to infer CME parameters at hot-Jupiter distances as the field of view covers up to 0.073\,au, that is 15.6\,R$_{\odot}$. The velocity of this CME has been determined with additional multi-point coronagraph views (by \emph{STEREO-Behind} and \emph{SOHO}) and forward-modeling of a 3D shape for the CME obtaining 1277$\pm$127\,km\,s$^{-1}$ \citep{mostl2014}. This specific CME serves as our medium CME case with a speed of 1300\,km\,s$^{-1}$. 

Figure~\ref{picture:cor_cme} also shows that a CME usually consists of three parts: the leading edge, which is a pile-up of plasma between the shock and the magnetic flux rope, the low-density void containing the magnetic flux rope, and the higher-density core, which is often associated with filament or prominence material erupting along with the CME \citep[e.g.,][]{mostl2009,vourlidas2013}. Note that the intensity observed in coronagraphic images is related to the integrated density along the line of sight from the observer through the coronal plasma, and higher intensity implies higher integrated density. The density profiles at a particular point in coronagraphic images are not available from the literature and their detailed derivation goes beyond the scope of the present work. Therefore, we determined empirically three different approximate CME density and velocity profiles that an exoplanet at 0.05\,au would experience if it orbited a Sun-like star.

Figure~\ref{picture:wind} shows the velocity and density profiles for the considered slow, medium, and fast CMEs, as a function of time since the CME eruption, in hours. The three different CMEs have different propagation velocities measured at the CME front, that is 3000, 1300, and 600\,km\,s$^{-1}$ for the fast, medium, and slow CMEs, respectively. We have also included a small linear deceleration to the velocity profiles for the fast and medium cases, because CMEs faster than the background wind decelerate, which results in a slower speed of the core compared to the leading edge. 

Consequently, the three CMEs have different durations at the planet location: 1.3, 3.0, and 6.4\,hr for the fast, medium, and slow CMEs, respectively. For every CME, the duration of each phase (leading edge, void, core) is one-third of the total duration of that CME. We have determined the duration of the CME at the planet position by assuming that it propagates with the given velocity and radial CME size of 20 solar radii over the planet. This simple approximation is valid for all three CME cases, and leads to the correct duration of the CMEs of the respective velocity at a given point in the coronagraphic images. We have checked our results for the CME velocities using movies of the solar CMEs that occurred on 2012 July 23  for the fast case \citep{liu2014}, on 2012 July 12 for the medium case, and on 2008 December 12 for the slow case \citep[e.g.,][]{mostl2014}. 

The first phase of each CME corresponds to the passage of the shock and its sheath region of elevated density, and is characterized by an increase in the plasma velocity by factors of 30, 13, and 6 for the fast, medium, and slow CMEs, respectively. For each CME, the density during the first phase increases 10 times compared to the background wind. During the second phase, the density is similar to that of the background wind, while in the third phase the density increases again by a factor of 5 compared to that of the background wind. 

Note that these values are estimates representative of the average CMEs observed erupting from the Sun. It has been shown by \citet{ontiveros2009} that faint structures in front of CME leading edges have densities up to a factor of 3 higher than those in the background corona, so a factor of 10 for the leading edge is the correct order of magnitude. The second phase has much lower densities than the first, as the magnetic flux rope is largely empty, though the dynamical pressure during the flux rope is still higher than that in the stationary case. During the CME, the plasma temperature is similar to that of the background wind. After the end of the third phase, the wind parameters return to their initial values. Note that our simulations did not take into account variations in the chemical composition of the wind.

We modeled the CME as a time-variable boundary condition, starting from the initial solution of the quasi-closed atmosphere given in \citet{Bisikalo-2013b-Rus}. Once the CME has crossed the region of space covered by the planet, the wind parameters were set equal to those of the stationary wind. For all simulations, we also assumed that the planet lies inside the boundaries of a CME along the whole duration of the CME.
\section{Results}\label{sec:cmeresult}
\begin{figure*}[h!]
\begin{center}
\includegraphics[width=\hsize,trim={0cm 1.5cm 0 0.90cm},clip]{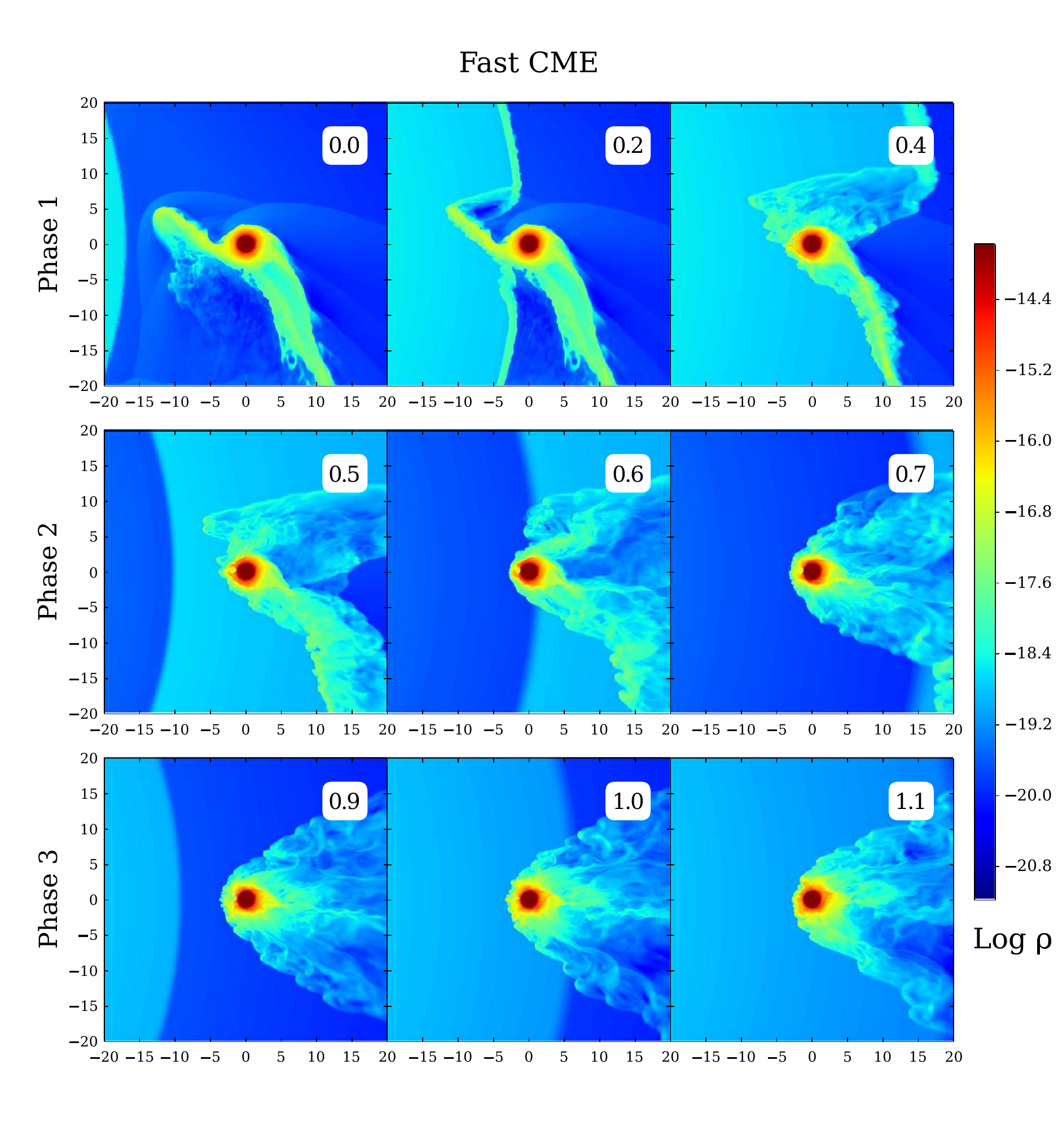}
\caption{Snapshots of the simulation along the planet orbital plane of the density distribution of the quasi-closed envelope subject to the fast CME. The time, in hours, is given in the top right corner of each panel. The star is located to the left. The axes are in units of \Rpl.}
\label{plotCollage} 
\end{center}
\end{figure*}

\begin{figure*}[h!]
\begin{center}
\includegraphics[width=\hsize,trim={0cm 1.5cm 0 0.86cm},clip]{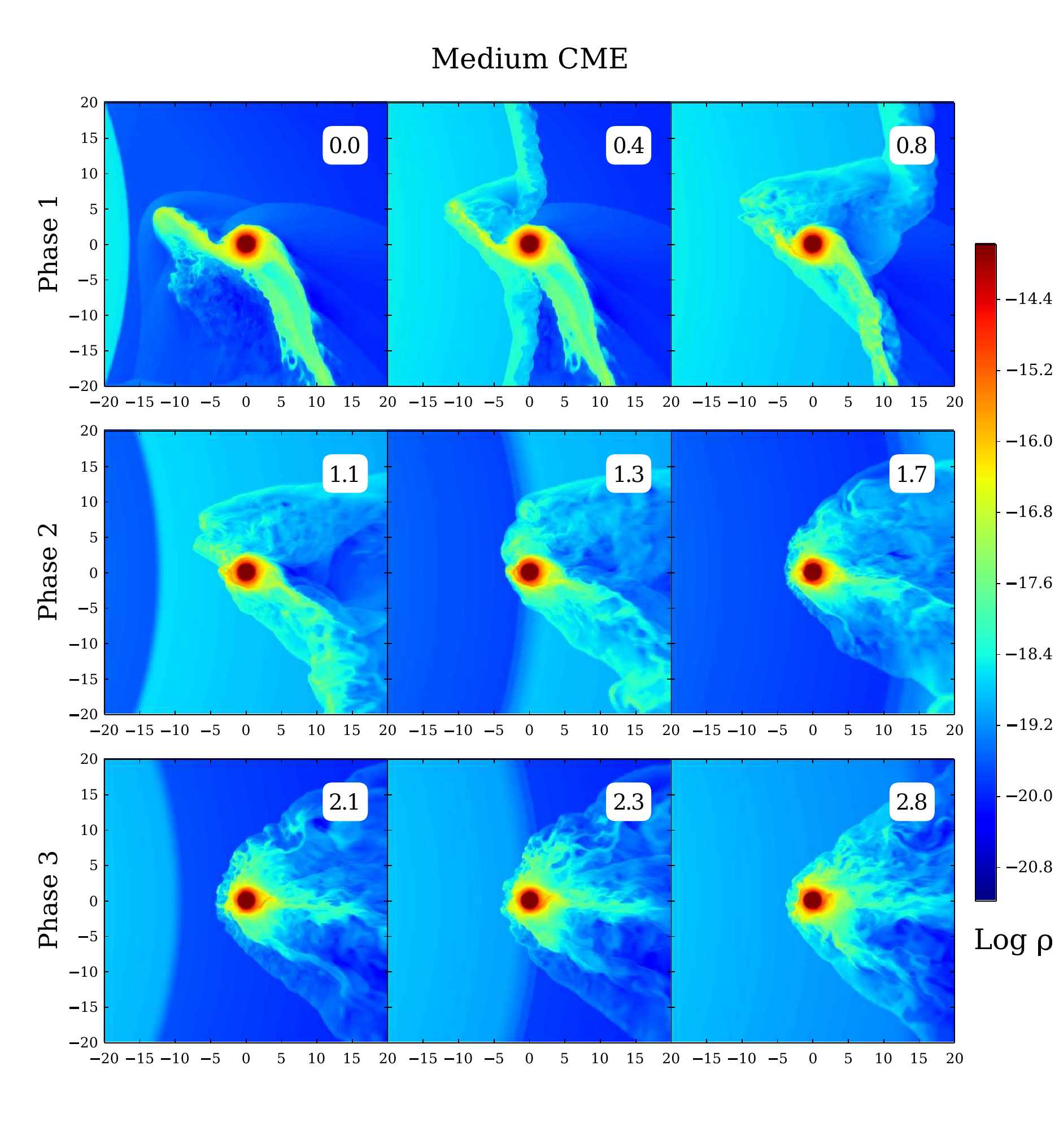}
\caption{Same as Fig.~\ref{plotCollage}, but for the medium CME.}
\label{plotCollage_medium} 
\end{center}
\end{figure*}

\begin{figure*}[h!]
\begin{center}
\includegraphics[width=\hsize,trim={0cm 1.5cm 0 0.86cm},clip]{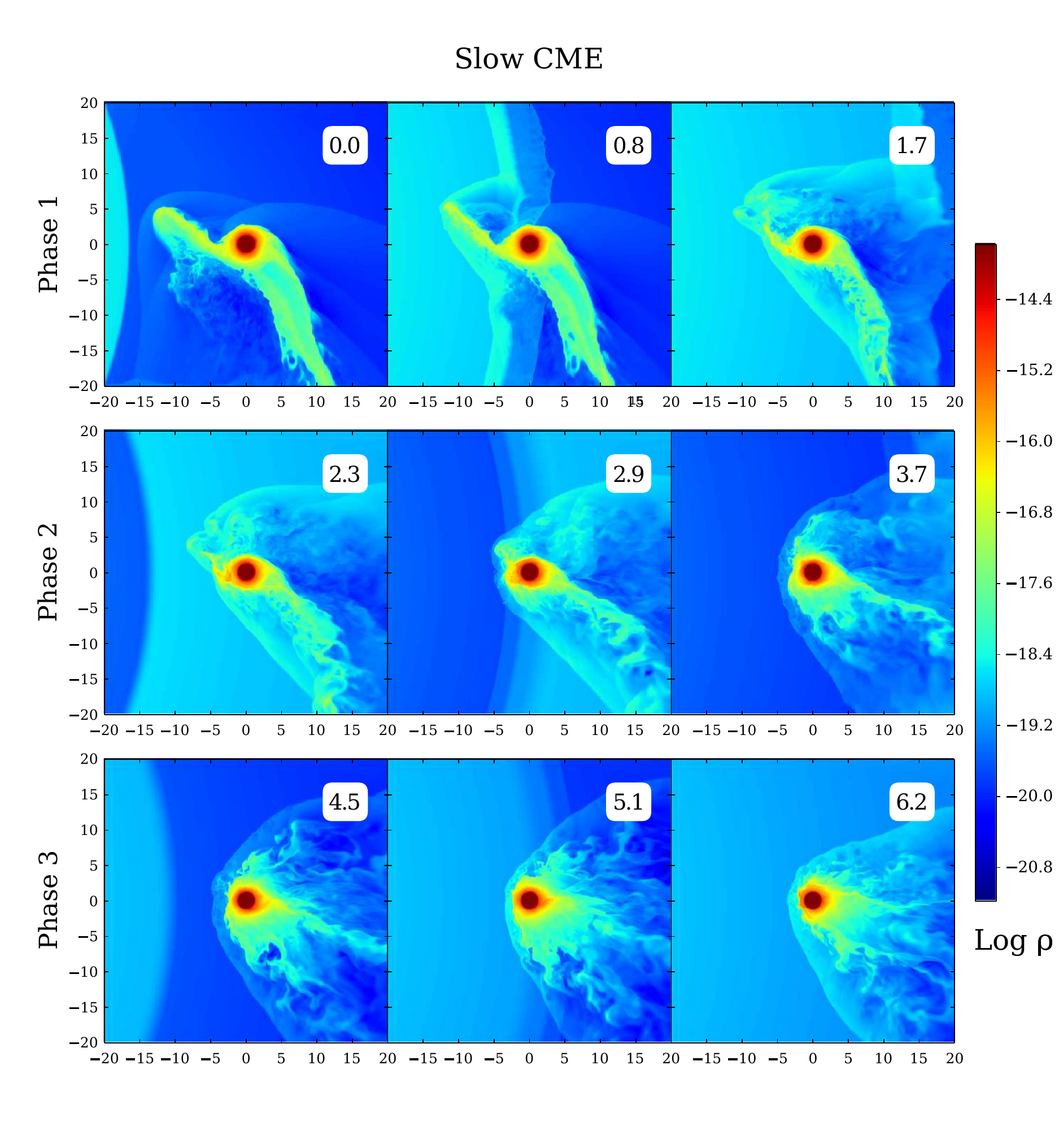}
\caption{Same as Fig.~\ref{plotCollage}, but for the slow CME.}
\label{plotCollage_slow} 
\end{center}
\end{figure*}

Figures~\ref{plotCollage}--\ref{plotCollage_slow} show the results obtained from modeling the interaction between the exosphere of HD 209458b and the fast, medium, and slow CMEs, respectively. Each panel shows the density distribution on the orbital plane of the system. Within each plot, each row corresponds to one of the three CME phases and the snapshots have been chosen to show the beginning, middle, and end of each phase of the simulation. Since the results are qualitatively similar for all three considered CMEs, here we only discuss in detail the case of the fast CME, but the following considerations are also valid for the medium and slow CMEs.

The density distributions displayed in the upper row of Figure~\ref{plotCollage} correspond to the first phase of the CME, i.e., the first rise in density shown in Figure~\ref{picture:wind}. The top left panel corresponds to the time before the arrival of the CME, hence to the case with the stationary wind  \citep{Bisikalo-2013b-Rus}. The top middle panel shows the arrival of the first front of the CME, which has a higher dynamical pressure compared to that of the stationary wind ($\rho_1 {v_1}^2/\rho_{st} {v_{st}}^2 \approx 9 \times 10^{3}$) and is therefore accompanied by the propagation of a shock across the computational domain. The collision of the front of the CME with the planetary envelope disrupts the outflow from the \Lp1 point (top right panel) and shifts the vortical wake of the stationary flow. 

The middle row of Figure~\ref{plotCollage} corresponds to the second phase, between the two density peaks shown in the bottom panel of Figure~\ref{picture:wind}. In this phase, the density drops by a factor of 10 relative to the previous phase, but its ram-pressure is still much higher than that of the stationary wind ($\rho_2 {v_2}^2/\rho_{st} {v_{st}}^2 \approx 8 \times 10^{2}$). This second phase of the CME completely destroys the two outflow structures emerging from \Lp1 and \Lp2. Toward the end of this second phase, the high dynamic pressure of the wind suppresses the outflow from the \Lp1 point, while the outflow from the \Lp2 point shifts to the direction opposite from that of the star.

The bottom row of Figure~\ref{plotCollage} shows the interaction between the remaining of the planetary exosphere and the third phase of the CME (i.e., second density peak in Figure~\ref{picture:wind}), where the dynamical pressure increases again to values similar to those of the first phase ($\rho_3 {v_3}^2/\rho_{st} {v_{st}}^2 \approx 3.6 \times 10^{3}$). During this phase, the solution remains very similar to the final one of the previous phase, despite the higher dynamical pressure, with the difference residing in the wind density.

\begin{figure}[h!]
\begin{center}
\includegraphics[width=\hsize,clip]{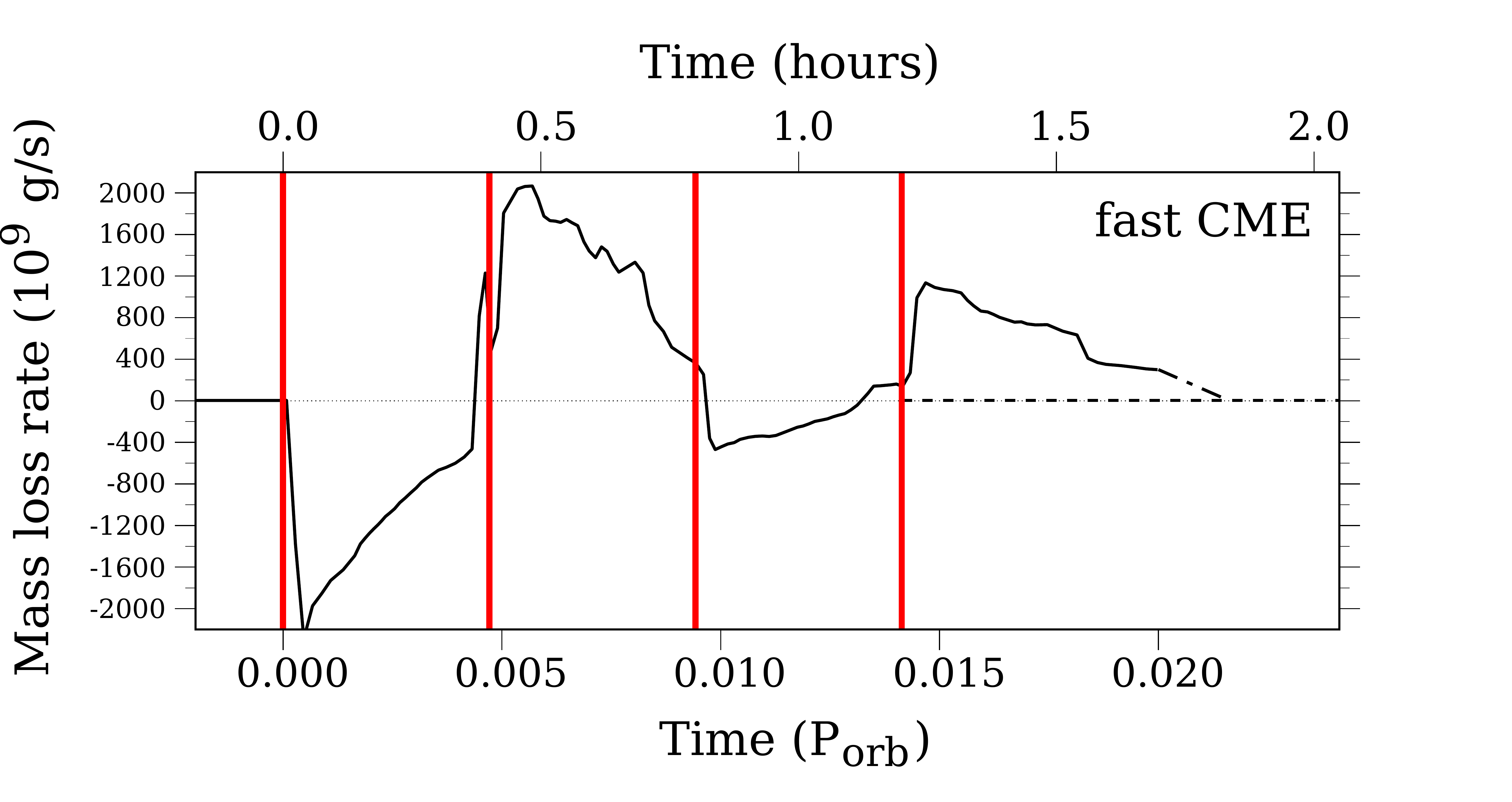}
\includegraphics[width=\hsize,clip]{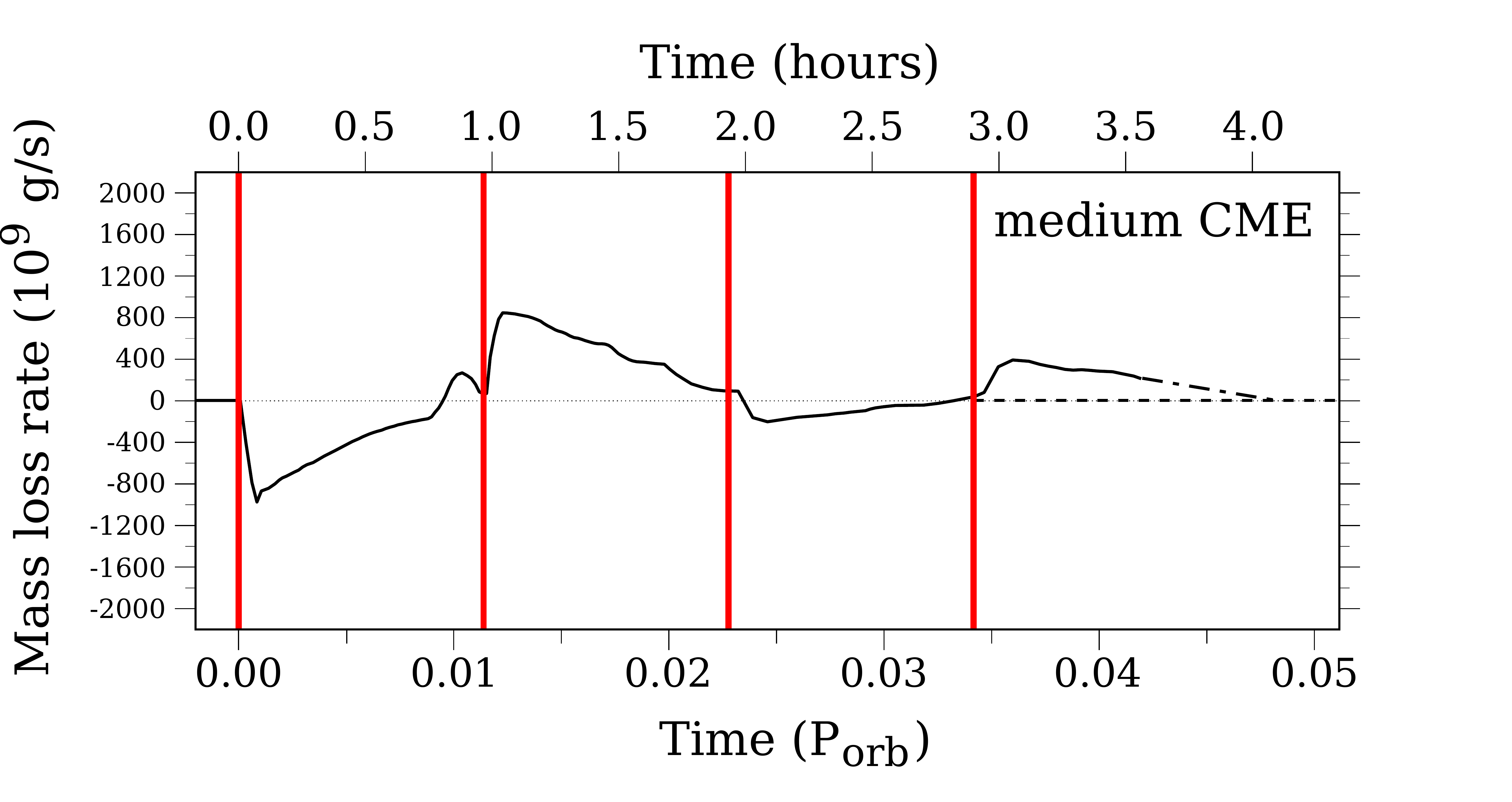}
\includegraphics[width=\hsize,clip]{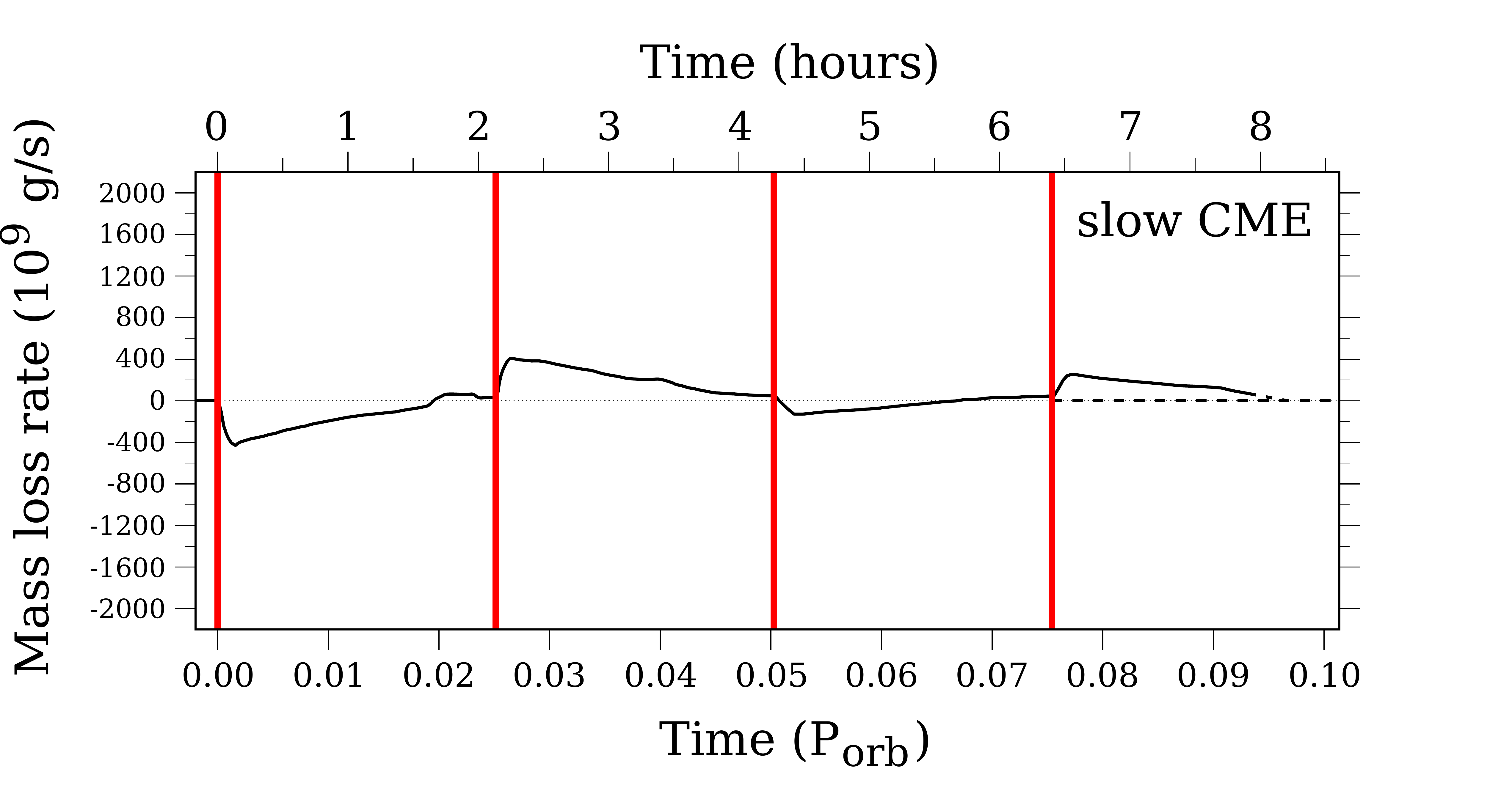}
\caption{Temporal variations of the mass-loss rate during the crossing of the fast (top), medium (middle), and slow (bottom) CME. The vertical red lines indicate the boundaries of the three phases of the CME described in Sect.~\ref{sec:cme-params}. The dashed-dotted line at the end of the simulation shows the extrapolation of the mass-loss rate toward the level obtained with the stationary wind (dotted line) of $3.0 \times 10^9$\,\gs. Note that the CMEs have different durations.}
\label{resultFast} 
\end{center}
\end{figure}

The similarity between Figures~\ref{plotCollage}--\ref{plotCollage_slow} is due to the fact that the duration of each CME multiplied by the propagation velocity of the CME is almost the same in all three cases of slow, medium, and fast CME. The difference between the results obtained for the three cases is mainly in the mass-loss rates, which are shown in Figure~\ref{resultFast} as a function of time. We remind the reader that in our simulations the mass-loss rates are calculated as the total mass crossing a box centered on the planet and being as large as $32\times32\times10$\,\Rpl. For this reason, the mass-loss rates can be negative when, for example, a CME approaches the planet (i.e., stellar wind material entering the box). The mass-loss rate returns to the values obtained for the stationary wind after the crossing of the fast, medium, and slow CMEs at 0.6, 1.2, and 2\,hr, respectively. 

%
\section{Discussion}\label{sec:discussion}
Table~\ref{tabular:results} lists the total mass-losses and average mass-loss rates obtained considering the three CMEs and compares it with the results of \citet{Cherenkov-et-al-15}. The mass-loss rate obtained with the stationary wind is $3.0 \times 10^9$\,\gs\ \citep{Cherenkov-14}. In Table~\ref{tabular:results}, the total duration is the sum of the CME duration and the time needed for the mass-loss rate to return to the initial value obtained with the stationary wind. The total mass-loss is derived by integrating over time, from the time of the CME hit until the time the mass-loss rate returns to the initial value, which is the mass-loss rates shown in Figure~\ref{resultFast}. The average mass-loss rate is the ratio of the total mass lost over the duration of the interaction with the CME, including the time required for the mass-loss rate to return to the starting value.

\begin{table*}[t!]
\caption{Duration and Mass-loss Obtained from Each Simulation and Compared to Those of \citet{Cherenkov-et-al-15}.}\label{tabular:results}
\begin{center}
\begin{tabular}{c|c|c|c|c}
\hline
\hline
CME & Fast & Medium & Slow & \citet{Cherenkov-et-al-15} \\
\hline
CME duration (hours)   & 1.3 & 3.0 & 6.4 & 42 \\
Total duration (hours) & 1.8 & 4.1 & 8.2 & 70 \\ 
\hline
Total mass loss ($10^{15}$\,g) & 1.5 & 1.0 & 1.2 & 10 \\ 
Average mass-loss rate ($10^9$\,\gs) & 226 & 71 & 39 & 42 \\
\hline
\end{tabular}
\vspace{2ex}

\raggedright Note. The total duration is the time needed for the mass-loss rate to return to
the initial value, while the average mass-loss rate is the ratio of the total mass
lost over the duration of the interaction with the CME.
\end{center}
\end{table*}

The last column of Table~\ref{tabular:results} shows, for comparison, the results obtained by \citet{Cherenkov-et-al-15}. It shows that there is a difference of about a factor of 10 in the total mass lost after the crossing of a CME, with the value of \citet{Cherenkov-et-al-15} being the higher one. This is because \citet{Cherenkov-et-al-15} did not consider that the duration of a CME changes along its propagation to 1\,au, where \citet{Cherenkov-et-al-15} took their CME parameters.

We obtain that the total mass lost is approximately the same for all three considered CMEs and is ${\approx}10^{15}$\,g. This result is somewhat expected because the duration of each CME is inversely proportional to its propagation velocity. We can therefore conclude that for CMEs falling withsin the range of parameters considered here, the total mass lost is approximately the same, though it may not be possible to extrapolate this result for other stronger/weaker CMEs because the interaction between the CME and the planetary exosphere is nonlinear.

Assuming a solar-like CME rate, we know from near-Earth solar wind observations that CMEs impact Earth about twice per month, or 23 times per year \citep{richardson2010}. This is not expected to be strongly different for the much closer hot-Jupiter orbit for two reasons. First, CMEs expand self similarly, so they maintain the same angular width (on average 60$^{\circ}$) while propagating away from the Sun, which does not change the impact rate. Second, for a 3.5\,day orbit, the planet moves for about 6$^{\circ}$--28$^{\circ}$ around the star during a CME impact, assuming that this impact lasts for 1.3 (fast CMEs) to 6.4\,hr (slow CMEs). As a result, the probability of a CME hit is about the same for a hot-Jupiter and a ``stationary'' planet, such as Earth.

Following the above considerations, we can derive the total mass lost by the hot-Jupiter as a result of CME impacts. Considering 23 CME hits in one year and that the mass lost by the planet in one CME crossing is about 10$^{15}$\,g, we find that the mass lost by the planet because of CME impacts in 1\,Gyr is about 2$\times$10$^{25}$\,g. Note that the mass of the planet (core$+$envelope) is about 10$^{30}$\,g. During the periods without CME hits, the planet loses mass because of the input from the high-energy stellar flux, with a mass-loss rate of about 3$\times$10$^9$\,g\,s$^{-1}$, which in 1\,Gyr implies a total mass-loss of about 9$\times$10$^{25}$\,g. It follows that the mass lost because of the high-energy stellar flux and because of CME encounters in 1 Gyr is about the same. Although we do not have any statistics for CMEs as a function of stellar age, it is plausible to believe that the CME rate decreases with stellar age in a way not too dissimilar from that of the high-energy stellar flux does. This implies that, as a rule of thumb, the similarity between the mass lost because of the high-energy stellar flux and the mass lost because of CME crossings could be the same during the time the star lies on the main-sequence.
\section{Conclusion}\label{sec:conclusion}
We investigated the influence of CME impacts on the gaseous envelope of the typical hot-Jupiter HD\,209458b. In particular, we considered three CMEs with different durations and propagation velocities. We found that the first encounter of the CME with the planetary atmosphere strips away the gravitationally unbound part of the quasi-closed envelope.

We obtained that for the three different CMEs, the planetary mass lost during the complete crossing of a CME is of the order of 10$^{15}$\,g, which becomes about 10$^{25}$\,g over the course of 1\,Gyr. This value is of the same order of magnitude as the mass lost by the planet in 1\,Gyr under the effect of the high-energy stellar flux and it is plausible to believe that this similarity can be extended across the time spent by the star on the main-sequence, though detailed calculation should be necessary to prove this for  faster and denser CMEs (probably the case for young stars) compared to what is considered here. This suggests that it is possible to account for CME impacts in atmospheric evolution calculations for hot-Jupiters by simply doubling the mass-loss rates obtained from energy-limited calculations \citep[e.g.,][]{erkaev2007,lopez2013}.

This work shows that CME encounters have a non-negligible impact on the overall mass lost by a Jupiter-mass planet in close orbit to a solar-like star. This implies that the effects of CMEs, and of space weather in general, should be taken into account in planetary evolution calculations, particularly for close-in planets.


%
\bigskip
This work is supported by the Russian Scientific Foundation and the Russian Foundation for Basic Research.. The results of this work were obtained using the computational resources of the MCC NRC Kurchatov Institute (http://computing.kiae.ru/). C.M. acknowledges support by the Austrian Science Fund (FWF): P26174-N27. The presented work has received funding from the European Union Seventh Framework Programme (FP7/20072013) under grant agreement No. 606692 [HELCATS]. We thank Ute Amerstorfer for useful discussions.
%
%

\end{document}